\documentclass[aps,preprint,showpacs,showkeys]{revtex4}%
\usepackage{amsmath,bm}
\usepackage{amsmath}
\usepackage{amsfonts}
\usepackage{amssymb}
\usepackage{graphics}  
\usepackage{graphicx}
\newcommand{\csch}{\textrm{ csch }}
\newcommand{\sech}{\textrm{ sech }}

\newcommand{\be}{\begin{eqnarray}}
\newcommand{\ee}{\end{eqnarray}}

\begin{document}
\date{\today }
\title{The Higgs oscillator on the hyperbolic plane and Light Front Holography      }
\author{A.\ Pallares-Rivera and M.\ Kirchbach}
\affiliation{$^2$Instituto de F\'{\i}sica, \\
         Universidad Aut\'onoma de San Luis Potos\'{\i},\\
         Av. Manuel Nava 6, San Luis Potos\'{\i}, S.L.P. 78290, M\'exico}


\begin{abstract} 
The Light Front Holographic (LFH) wave equation, which is the conformal scalar equation on the plane, is revisited
from the perspective of the supersymmetric quantum mechanics, and 
attention is drawn to the fact that it naturally emerges in the small hyperbolic angle approximation 
to the ``curved'' Higgs oscillator on the hyperbolic plane, i.e. on the upper part of the two-dimensional hyperboloid of 
two sheets, {\bf H}$_{+R}^2$, a space of constant negative curvature.
Such occurs because the particle dynamics  under consideration reduces to the one dimensional 
Schr\"odinger equation with the second hyperbolic P\"oschl-Teller potential, whose flat-space (small-angle)
limit reduces to the conformally invariant inverse square distance plus harmonic oscillator interaction,
on which  LFH is based. In consequence, energies and wave functions of the  LFH spectrum can be approached by the 
solutions of the Higgs oscillator on the hyperbolic plane in employing its curvature and the potential strength as 
fitting parameters.  Also the proton electric charge form factor is well reproduced within this scheme by means of a Fourier-Helgason hyperbolic wave transform of the charge density. In conclusion, in the small angle approximation, the Higgs oscillator on 
{\bf H}$_{+R}^2$ is demonstrated to satisfactory parallel  essential outcomes of the Light Front Holographic QCD. 
The findings are suggestive of  associating the {\bf H}$_{+R}^2$  curvature with a second scale in LFH, which then could be employed in the definition of a chemical potential.
\end{abstract} 


\maketitle

\section{Introduction to the LFH wave equation. Goals and motivation.}

Supersymmetric quantum mechanics  has exercised over the years  notable influence on various fields of spectroscopic studies by providing an 
efficient machinery  in handling a large variety of interactions in complex systems by means of exactly solvable potentials \cite{CooperKhare}.
Specifically in particle physics, one repeatedly faces situations in which most complicated field-theoretical considerations  amount to known Sturm-Liouville problems,
a prominent example being the Light-Front Holographic QCD \cite{Review}--\cite{BdTD}.
This approach is derived from the first-principle motivated gauge-gravity duality concept and is 
quite successful in describing a broad range of hadron properties. 
At its root one finds two effective one-dimensional Schr\"odinger equations with an inverse square distance plus harmonic oscillator potential, which defines
the conformal scalar equations on the plane according to, 
\begin{eqnarray}
\left(-\frac{{\mathrm d}^2}{{\mathrm d}\zeta^2}
+\frac{\nu^2 -\frac{1}{4}}{\zeta^2 }+\kappa^4\zeta^2  +c_+^\nu\right)\Psi^{n\nu}_+(\zeta)=E^2\Psi^{n\nu}_+(\zeta),\,\,
c^\nu_+ = 2\kappa^2(\nu+1),&&
\label{Gl1}\\
\left(-\frac{{\mathrm d}^2}{{\mathrm d}\zeta^2} +\frac{(\nu+1)^2 -\frac{1}{4}}{\zeta^2 }+\kappa^4\zeta^2  +c_-^\nu \right)
\Psi^{n(\nu+1)}_-(\zeta)=E^2
\Psi^{n (\nu+1)}_-(\zeta),\,\, c_-^\nu=2\kappa^2\nu,&&
\label{Gl1_n}
\end{eqnarray}
with $1/\kappa$ being an external lengths scale. 
The equations (\ref{Gl1})-(\ref{Gl1_n}) can be viewed either as  one-dimensional Klein-Gordon equations, or equivalently, 
as one-dimensional Schr\"odinger equations under the identifications,  
\begin{equation}
E^2=\frac{2\mu c^2E^{Schr}}{\hbar^2c^2},\quad \kappa^4=\frac{\mu^2\omega^2}{\hbar^2},\quad \left[ E^2\right]={\mbox{fm}}^{-2},\quad E^2={\mathcal M}^2,
\label{unidades}
\end{equation}
where $E^{Schr}$ is the  energy in MeV of a particle of mass $\mu$ in the stationary Schr\"odinger equation or, the reduced mass 
of a two-body system, such  as quark--diquark (q-qq), with $\zeta$ being associated with a relative distance. 
In  introducing the operators, $B^+_\nu$ and $B_\nu$, laddering back and forth between 
$\Psi_-^{n(\nu+1)}(\zeta)$ and $\Psi_+^{n\nu}(\zeta)$  according to,  
\begin{eqnarray}
B^+_\nu\Psi_{-}^{n(\nu+1)}(\zeta)&=&E\Psi_{+}^{n\nu }(\zeta),
\quad B^+_\nu=
-\frac{ {\mathrm d} }{{\mathrm d}\zeta }  -
\frac{\nu +\frac{1}{2}}{\zeta} -\kappa^2 \zeta,
\nonumber\\
B_\nu\Psi_{+}^{n\nu}(\zeta)&=& E\Psi_{-}^{n(\nu+1) }(\zeta),\quad B_\nu =\frac{ {\mathrm d} }{{\mathrm d}\zeta }  -
\frac{\nu +\frac{1}{2}}{\zeta} -\kappa^2 \zeta , \quad
\label{coupled}
\end{eqnarray}
one finds a Dirac-like   factorization of the equations (\ref{Gl1}) and (\ref{Gl1_n}),
\begin{eqnarray}
B^+_\nu   B_\nu \Psi_+^{n\nu}=E^2\Psi_+^{n\nu}, &\quad&   B_\nu B^+_\nu \Psi_-^{n (\nu+1)}=E^2\Psi_-^{n(\nu+1)},
\end{eqnarray}
admittedly with vectorial rather than scalar inverse-distance plus linear potentials.
The explicit solutions to (\ref{Gl1}) and (\ref{Gl1_n}) read,
\begin{eqnarray}
\Psi^{n\nu}_+(\zeta) &=&N_{n\nu}(\kappa^2\zeta^2)^{\frac{\nu}{2}+\frac{1}{4}}e^{-\frac{\kappa^2\zeta^2}{2}}L_n^\nu (\kappa^2\zeta^2),
\, L_n^\nu (\kappa^2\zeta^2)\sim \, _1F_1(-n,\nu +1, \kappa^2\zeta^2),
\nonumber\\
\Psi^{n(\nu+1)}_-(\zeta)&=&{\bar N}_{n\nu}(\kappa^2\zeta^2)^{\frac{\nu +1}{2}+\frac{1}{4}}e^{-\frac{\kappa^2\zeta^2}{2}}
L_n^{\nu+1} (\kappa^2\zeta^2), \, E^2=4\hbar^2c^2 \kappa^2\left(n+\nu +1\right)\nonumber\\
&=&4\hbar^2c^2 \kappa^2\left(n+\frac{\nu +1}{2}\right)  + c^\nu_+
=4\hbar^2c^2 \kappa^2\left(n+\frac{\nu +2}{2}\right)  + c^\nu_-,
\label{Gl2}
\end{eqnarray}
with $c^\nu_+$ from (\ref{Gl1}) and $c_-^\nu$ from (\ref{Gl1_n}).
Here $L_n^\nu$ are the generalized Laguerre polynomials, $N_{n\nu}$, and ${\bar N}_{n\nu}$  are normalization constants, and $_1F_1$ stands for the confluent hypergeometric function.
In being  distinct by $\Delta \nu=1$,  $\Psi^{n\nu}_+(\zeta)$, and $\Psi^{n(\nu+1)}_-(\zeta)$ are given within the LFH method under discussion
the interpretation of a large and a small component of a Dirac spinor, thus allowing to correct for the scalar nature of
eqs.~(\ref{Gl1})--(\ref{Gl1_n}).  

The first  goal of the present study is to comment on some specific  supersymmetric quantum mechanical aspects of the above 
equations (\ref{Gl1}) and (\ref{Gl1_n}) and explore consequences. 
The pair of  wave functions, $\Psi^{n\nu}_+(\zeta)$, and $\Psi_-^{n(\nu +1)}(\zeta)$, in reality  describe states 
belonging  to either  even or odd principal quantum numbers, $N_+=(2n+\nu) $, and $N_-=(2n+\nu +1)$, 
of the three dimensional harmonic oscillator,  distinct by one unit. In addition, they 
belong to  a pair of supersymmetric partner spectra, and 
the corresponding SUSY-QM equations satisfied by them predict distinct energy excitations, 
$E_+=4\hbar^2c^2\kappa^2n$, versus  $E_-=4\hbar^2 \kappa^2 (n+1)$, respectively.
This energy separation has been removed in the above LFH equations (\ref{Gl1}) and (\ref{Gl1_n})
through shifts  of  the respective SUSY-QM Hamiltonians by the  additive constants $2c_+^\nu$, and $2c_-^{\nu}$,
respectively. 
Gaining a deeper  insight into the nature of the  $\Psi^{n\nu }$--$\Psi^{n(\nu+1)}$  wave functions
would allow  to understand  as two what extent the consideration of
 such pairs as Dirac spinor components is amenable to generalizations  to other potentials.

The next goal of the present study is to show that ~(\ref{Gl1})-(\ref{Gl1_n}) represent the decreasing curvature limit of 
quantum motion on the hyperbolic plane within the Higgs oscillator potential and to explore consequences.
The paper is structured as follows. The next section is devoted to the SUSY-QM aspects of the equations (\ref{Gl1})-(\ref{Gl1_n}).
In section 3 we briefly highlight the basics of quantum motion on the hyperbolic plane, define
the Higgs oscillator potential problem there, and show that it is equivalent to the one dimensional Schr\"odinger equation with
the generalized hyperbolic P\"oschl-Teller potential \cite{CooperKhare}. In section 4 we take the small hyperbolic angle limit 
of the latter equation and its solutions and reveal their equivalence to the LFH wave equation and its solutions.
We furthermore calculate the proton electric charge form factor and draw conclusions on the relevance of the conformal symmetry in the 
two extreme regimes of QCD, the infrared and the ultraviolet. The paper closes with brief conclusions and has one Appendix.

\section{The SUSY-QM aspects of Light-Front Holography and conformal symmetry}
We first notice that, modulo the additive constants, $c_\nu^+$, and $c_\nu^-$, the equations (\ref{Gl1}), (\ref{Gl1_n}) are known from the 
\underline{S}uper\underline{S}ymmetric \underline{Q}uantum \underline{M}echanics  (SUSY-QM) of the three dimesnional (3D) 
harmonic oscillator in the presence of a centrifugal barrier \cite{CooperKhare}. 
In using $\hbar=1$, $2\mu=1$ units, and introducing the operators $A_\nu$ and $A^+_\nu$,
as
\begin{eqnarray}
A^+_\nu=-\frac{{\mathrm d}}{{\mathrm d} \zeta } +{\mathcal W}_\nu(\zeta), &\quad& 
A_\nu=\frac{{\mathrm d}}{{\mathrm d} \zeta } +{\mathcal W}_\nu(\zeta), \quad
{\mathcal W}_\nu (\zeta)= -\frac{\nu +\frac{1}{2}}{\zeta} +\kappa^2\zeta,
\label{factoriz}
\end{eqnarray}
where ${\mathcal W}_\nu (\zeta)$ is the superpotential,
it is straightforward to calculate that the  $n=0$ function,  $\Psi^{01}_+(\zeta)$ in  (\ref{Gl2}) 
is  nullified by $A_{1}$, 
\begin{equation}
A_{1}\Psi^{01}_+(\zeta)=0,
\label{grstts}
\end{equation}
meaning that it refers to a genuine SUSY-QM ground state.
This ground state is vanishing in the $\zeta\to \pm \infty$ limit, as it should be.
Therefore, the  LFH ground state wave function  in (\ref{Gl1}) is  consistent with the SUSY-QM definition of a ground state
in terms of the superpotential as, 
\begin{equation}
\Psi^{01}_+ (\zeta)\sim \exp \left(-\int_0^\zeta  {\mathcal W}_1 (y ){\mathrm d}y \right)\sim \zeta^{\nu+\frac{1}{2}}
 e^{-\frac{1}{2}\kappa^2\zeta^2}.
\label{SUSU-gst}
\end{equation} 
In now introducing the  two  SUSY-QM supercharges (for any general $ \nu\geq 1 $)
$Q_\nu$, and $Q^+_\nu$, 
\begin{eqnarray}
Q_\nu=\left(
\begin{array}{cc}
0&0\\
A_\nu&0\\
\end{array}
\right), &\quad &
Q^+_\nu=\left(
\begin{array}{cc}
0&A^+_\nu\\
0&0\\
\end{array}
\right), 
\label{supercharges}
\end{eqnarray} 
satisfying the $sl(1/1)$ superalgebra, one defines the standard matrix SUSY-QM Hamiltonian as,
\begin{eqnarray}
H=\lbrace Q_\nu,Q_\nu^+\rbrace =
\left(
\begin{array}{cc}
 H_+&0\\
 0  &H_-
\end{array}
\right)
 =
\left( 
\begin{array}{cc}
A^+_\nu A_\nu &0\\
0            &A_\nu A^+_\nu 
\end{array}
\right),&&\nonumber\\ 
H\Psi^{n\nu} (\zeta) = \left(E_S^{n\nu}\right)^2 \Psi^{n\nu} (\zeta), \,\,  \Psi^{n\nu} (\zeta)=
\left( 
\begin{array}{c}
\Psi_+^{n\nu}(\zeta)\\
\Psi_-^{(n-1) (\nu+1)}(\zeta)
\end{array}
\right), \,\, \left( E^{n\nu}_S\right)^2=4\kappa^2 n.&&
\label{suprch}
\end{eqnarray}
In terms of $A_\nu$, and $A^+_\nu$, the superpotential ${\mathcal W}_\nu(\zeta)$, and its first derivative,
${\mathcal W}_\nu^\prime(\zeta)$,  
the wave functions $\Psi_+^{n\nu}(\zeta)$ and $\Psi_-^{n(\nu+1)}(\zeta )$ satisfy the following ``factorized'' 
wave equations \cite{CooperKhare}:
\begin{eqnarray}
&&A^+_\nu A_\nu \Psi_+^{n\nu}(\zeta)=\left( -\frac{{\mathrm d}^2}{{\mathrm d}\zeta^2}
+{\mathcal W}^2_\nu (\zeta) -{\mathcal W}_\nu^\prime(\zeta)
\right) \Psi_+^{n\nu}(\zeta)
=\left(E_S^{n\nu}\right)^2 \Psi^{n\nu}_+(\zeta),\nonumber\\
&&A_\nu A^+_\nu \Psi_-^{n(\nu+1)}(\zeta)
=
\left(-\frac{{\mathrm d}^2}{{\mathrm d}\zeta^2} + {\mathcal W}_\nu ^2(\zeta) +{\mathcal W}_\nu ^\prime (\zeta) \right)
\Psi^{n(\nu+1)}_-(\zeta)\nonumber\\
&&=\left( E_S^{n(\nu+1}\right)^2 \Psi^{n(\nu+1)}_-(\zeta), \,\,
\left(E_S^{n\nu}\right) ^2  =4\kappa^2 n, \, \left(E_S^{n(\nu+1)}\right) ^2=4\kappa^2(n+1). 
\label{mdt1}
\end{eqnarray}
The latter show that the first excited  level in $H_+$, with $n=1$, and  $\nu=1$,
has same energy as the ground state  of $H_-$, with $n=0$ and $\nu=2$. 
These two states are supersymmetric partners and are related by $A^+_1\Psi_-^{02}=\Psi_+^{11}$, more general
$A^+_\nu\Psi_-^{n(\nu+1)}=\Psi^{(n+1)\nu }_+$, holds valid.
The equations (\ref{mdt1}) can be cast into the form of the LFH equations (\ref{Gl1})-(\ref{Gl1_n})
 upon introducing the shifted partner Hamlitonians, ${\bar H}_+=\left({ H}_+ +2c_\nu^+\right)$, and 
${\bar H}_-=\left({H}_- +2c_\nu^-\right)$, as
\begin{eqnarray}
{\bar H}_+\Psi_+^{n(\nu)}(\zeta)
&=&\left(-\frac{{\mathrm d}^2}{{\mathrm d}\zeta^2}
+\frac{\nu^2 -\frac{1}{4}}{\zeta^2 }
+\kappa^4 \zeta^2 +c_\nu^+\right)\Psi_+^{n\nu}(\zeta)=E\Psi_+^{n\nu}(\zeta), 
\label{mtd2}\\
{\bar H}_-\Psi_-^{n(\nu+1)}(\zeta)
&=&\left(-\frac{{\mathrm d}^2}{{\mathrm d}\zeta^2} +
\frac{(\nu+1)^2 -\frac{1}{4}}{\zeta^2 }+\kappa^4\zeta^2 +c^\nu_- \right)
\Psi^{n(\nu+1)}_- (\zeta)\nonumber\\
&=&E\Psi^{n(\nu+1)}_-(\zeta).\label{hmenshft}
\end{eqnarray}
The LFH energy, $E$,  now expresses in terms of the SUSY-QM energies in (\ref{suprch}) as,
\begin{eqnarray}
E=4\kappa^2 \left( n+\nu +1 \right)=\left( E_S^{n\nu}\right) ^2 + 2c^\nu_+&=& 
\left( E_S^{n(\nu+1)}\right) ^2 +2c^\nu_- .
\label{SUSY_sols}
 \end{eqnarray}
In effect, in LFH the  former degenerate  SUSY partner states,
$\Psi_+^{n\nu}(\zeta)$ and $\Psi_-^{(n-1)(\nu+1)}(\zeta)$, 
acquire the energies,
$4\kappa^2 (n+\nu +1)$, and $4\kappa^2 (n+\nu )$ and appear split by the constant gap of $4\kappa^2$. 
In this way, the solutions to the  equations (\ref{Gl1})-(\ref{Gl1_n}) have been identified as
belonging to supersymmetric partner spectra, though the SUSY-QM degeneracies have then been removed  by the shifted Hamiltonians in 
(\ref{mtd2})--(\ref{hmenshft}) in favor of the ${\Big(} \Psi^{n\nu}$--$\Psi^{n (\nu+1)}{\Big)}$ degeneracy,  with the aim to 
interpret these functions as the large and small Dirac spinor components, respectively. That such a shift amounts to 
universal ${\Big(}\Psi^{n\nu}$--$\Psi^{n (\nu+1)}{\Big)}$ degeneracies for any $n$ and $\nu$ is inherent to the harmonic oscillator potential
and happens by virtue of the equidistant spacings between  its levels. For a general potential of non-equidistant level 
splittings, the universal degeneracy of states of equal nodes and angular momenta distinct by one unit will be  as a rule  beyond reach.  
Finally, the $B_\nu$ and $B_\nu^+$ operators in (\ref{coupled}) are  not useful as SUSY-QM factorizations  
because the  ground state defined as $B_\nu\phi_0=0$ is obtained as  $\phi_0\sim e^{\frac{\kappa^2\zeta^2}{2}}$ and is not vanishing at infinity, 
due to the negative sign of the $\kappa^2\zeta$ term.

{}For  $\nu^2 >\frac{1}{2},$ and $\kappa^4> 0$,  (\ref{Gl1})-(\ref{Gl2}) have the remarkable property to describe  propagation on the real line of a conformal particle in flat space-time and  have found application in the context of AdS$_2$/CFT$_1$ correspondence  \cite{GdTSBJ}, \cite{Barbon}, \cite{Leiva}, \cite{Karsch}, \cite{AW}, \cite{BdTD},
where one expects duality between a  string theory a  AdS$_2\times$ S$^2$ background and  a conformal theory on the boundary.
Indeed, the shifted SUSY-QM Hamiltonian, ${\bar H}_+=\left({ H}_++2c_\nu^+\right)$, in eq.~(\ref{mtd2}), again in units of $\hbar=1$, $2\mu=1$ for convenience, can be cast in the form of a linear combination of elements of a properly designed dynamical conformal $so(2,1)$ algebra,
according to \cite{AW}
\begin{eqnarray}
{\bar H}_+={ H}_++2c_\nu^+ &=&2\left(J_+ + \kappa^4J_-\right), \nonumber\\
J_-= \frac{1}{2}\zeta^2, &\quad& J_+= -\frac{1}{2}\left(\frac{
{\mathrm d}^2}{{\mathrm d} \zeta ^2} -
\frac{\nu^2 -\frac{1}{4}}{ \zeta^2}\right), \nonumber\\
\left[ J_+,J_-\right]=-2D_0, &\quad& \left[ D_0,J_\pm \right]=\mp J_\pm, \quad D_0=\frac{1}{4}\left( \zeta \, \frac{{\mathrm d}}{{\mathrm d}\, \zeta} +
\frac{{\mathrm d}}{{\mathrm d}\zeta} \zeta  \right).
\label{conf_alg}
\end{eqnarray}
A recent achievement in the field concerns the derivation in \cite{BdTD} of the $J_-$ related oscillator potential from the 
conformal action upon generalizing the Hamiltonian to a translation operator of an adequate  time variable.
In the next section we turn  to our second goal, which is to relate LFH to perturbed motion on the hyperbolic plane.

\section{Quantum motion on the hyperbolic plane}
The hyperbolic plane, {\bf H}$_{+R}^2$ is the upper part of a two-dimensional  hyperboloid of
two sheets, 
\begin{equation}
{\mathbf H}^2_{+R}:\quad x_1^2+x_2^2-x_0^2=-R^2,
\label{hypplane}
\end{equation}
where  $\left( -1/R^2\right)$ is the constant negative curvature of the surface under consideration.
The isometry algebra of {\bf H}$_{+R}^2$ is $so(1,2)$ and thereby the Lorentz group in $(1+2)$ dimensions.
In comparison, the conformal  $so(2,1)$ underlying LFH is  
a remnant of the conformal $so(2,4)$ algebra of the Minkowski space-time,
and acts as  isometry algebra of an AdS$_2$ space 
represented by the two-dimensional hyperboloid of one sheet. We shall comment on this point at a due place below. 
The  {\bf H}$^2_{+R}$ geometry we are interested in here,  derives its importance from the light cone,
\begin{eqnarray}
x_1^2+x_2^2+x_3^2-x_0^2&=&0,
\end{eqnarray}
restricted to
\begin{equation}
x_1^2+x_2^2-x_0^2 =-R^2=-x_3^2.
\label{H2pcs}
\end{equation}
In this section we shall highlight the essentials of quantum dynamics on the non-compact surface in (\ref{hypplane}).

\subsection{Free quantum motion on  $H^2_{+R}$ }
 In global coordinates,  the hyperbolic plane in (\ref{hypplane})  is parametrized as  (we closely follow presentation in 
\cite{cocoyoc} and references therein)
\begin{eqnarray}
x_1=R\sinh\rho \cos\varphi, &\quad& x_2=R\sinh\rho   \sin\varphi, \quad x_0=R\cosh \eta,\nonumber\\
{\mathcal C}(\rho, \varphi)  &=&\frac{1}{\sinh \rho }\frac{\partial }{\partial \rho }\sinh \rho \frac{\partial }{\partial \rho } +
\frac{\frac{\partial^2}{\partial \varphi^2}}{\sinh^2 \rho},
\label{ET}
\end{eqnarray}
with ${\mathcal C}(\rho,\varphi)$ standing for the geometric  $so(1,2)$ Casimir operator.
The quantum mechanical free motion is described
in terms of the  eigenvalue problem of ${\mathcal C}(\rho,\varphi)$, as
\begin{eqnarray}
-\frac{\hbar^2 }{2\mu R^2}{\mathcal C}(\rho,\varphi)\, Y_\ell^{m}(\rho, \varphi)&=&E^{so(1,2)}_\ell Y_\ell ^{m} (\rho ,\varphi), 
\quad Y_\ell ^{m}(\rho, \varphi)=P_\ell ^{m}(\cosh \rho ) e^{im\varphi},\nonumber\\
 E^{so(1,2)}_\ell &=&-\frac{\hbar^2}{2\mu R^2}\ell(\ell +1),
\label{ET1}
\end{eqnarray}
where $Y_\ell^m(\rho, \varphi)$ are the pseudo-spherical harmonics. 
\subsection{Equivalent description in terms of  the Schr\"odinger equation with the Eckart potential}
A suitable variable change,
\begin{eqnarray}
Y_\ell^m(\rho, \varphi)={P}_\ell^m(\cosh\rho)e^{im\varphi}=\frac{U_n ^m(\rho)}{\sqrt{\sinh\rho}}e^{im\varphi },  
\label{ET2}
\end{eqnarray}
(with $n$ defined in (\ref{wafuSch}) below)
converts the free motion on the hyperbolic plane, into an 1D Schr\"odinger equation with
the  Eckart potential, $\frac{a(a -1)}{\sinh^2\rho} $ \cite{CooperKhare}, \cite{Manning_Rosen}, and with $a$ chosen as  $a=|m|+1/2$, one finds
\begin{eqnarray}
{\mathcal H}(\rho,\varphi ) U_n ^m(\rho)e^{im\varphi}  =\left(E^{Schr}_n+\frac{\hbar^2}{8\mu R^2}\right)U_n^m(\rho)e^{im\varphi}
=  E_\ell^{so(1,2)}U_n^m(\rho)e^{im\varphi},&& \nonumber\\
{\mathcal H}(\rho,\varphi) = -\frac{\hbar^2}{2\mu R^2}\frac{1}{\sqrt{\sinh\rho}}{\mathcal C}(\rho,\varphi )\sqrt{\sinh\rho}=
-\frac{\hbar^2}{2\mu R^2}\frac{\partial^2}{\partial\rho^2}&&  \nonumber\\
+\frac{\hbar^2}{2\mu R^2}\frac{a(a -1)}{\sinh^2\rho }+\frac{\hbar^2}{8\mu R^2},\quad a =|m|+\frac{1}{2}.&&
\label{Eckart}
\end{eqnarray}
The energies  in (\ref{Eckart}) express in terms of the potential parameter $a$ in (\ref{Eckart}) as
\begin{eqnarray}
E^{Schr}_n+\frac{\hbar^2}{8\mu R^2}=-\frac{\hbar^2}{2\mu R^2}\left(a +n\right)^2+\frac{\hbar^2}{8\mu R^2}&=&
-\frac{\hbar^2}{2\mu R^2}\left(|m|+ n +\frac{1}{2}\right)^2 +\frac{\hbar^2}{8\mu R^2}\nonumber\\
=-\frac{\hbar ^2}{2\mu R^ 2}(|m|+n)(|m|+n+1) &=&E_\ell^{so(1,2)}.
\label{Eckrt_enrgs}
\end{eqnarray}
In view of (\ref{ET1}), the latter equation implies,
\begin{equation}
\ell =|m|+n, \quad  |m|\in [0,\ell].
\label{Eckart_ell}
\end{equation}
The wave functions are determined by Jacobi polynomials, $P_n^{\alpha,\beta}$, with parameters depending on their degree, $n$, as
\begin{equation}
U_n^m (\rho)= \sinh^{n+a}\rho P_n^{-n-a, -n-a }(\coth \rho)=\sinh^{|m|+n+\frac{1}{2}}\rho P_n^{-|m|-n-\frac{1}{2}, -|m|-n-\frac{1}{2}}(\coth\rho).
\label{wafuSch}
\end{equation}
In terms of the quantum numbers in (\ref{Eckart_ell}),  the wave functions take their  final forms as,
\begin{eqnarray}
U_n^m (\rho)= \sinh^{\frac{1}{2}}\rho P_\ell^{|m|}(\cosh\rho)&=&\sinh^{\frac{1}{2}}\rho \sinh^{|m|+n}\rho 
P_n^{-|m|-n-\frac{1}{2}, -|m|-n-\frac{1}{2}}(\coth\rho)\nonumber\\
&=&\sinh^{\frac{1}{2}}\rho \sinh^{\ell }\rho P_{\ell -|m|}^{-\left(\ell +\frac{1}{2}\right), -\left(\ell +\frac{1}{2}\right)}(\coth\rho).
\end{eqnarray}
This  equation by itself reproduces the following relationship between the associated Legendre functions and the Jacobi polynomials,
\begin{eqnarray}
P_\ell^{|m|}(\cosh\rho)=\sinh^{-\alpha_\ell -\frac{1}{2}}\rho P_{n}^{\alpha_\ell,\alpha_\ell}(\coth\rho),\quad
\alpha_\ell =-\ell -\frac{1}{2}, \quad n=\ell -|m|.&&
\label{Leg_Jac}
\end{eqnarray}
The latter equality means that if one were to prefer  pseudo-spherical harmonics to be expressed in terms of the Jacobi polynomials 
in place of the  associated Legendre functions of common use, the degree of the Jacobi polynomial has to be taken according to
(\ref{Leg_Jac}), (\ref{Eckart_ell}).
In consequence, the conclusion can be drawn that the isometry algebra of the surface on which the free motion takes place, acts
as a symmetry algebra of the potential (the one of Eckart in this case) appearing in the equivalent Schr\"odinger equation (\ref{Eckart}).

\subsection{The Higgs oscillator on H$^2_{+R}$ }
Now we shall introduce a perturbation of the free motion on {\bf H}$^2_{+R}$ in (\ref{ET1}) by an oscillator interaction.
The general definition of an oscillator on a curved surface, referred to as Higgs oscillator \cite{Higgs},
is given in terms of the square tangent to a geodesic.
Specifically on the hyperboloid  under consideration it is introduced as
\begin{eqnarray}
V_{\mbox{Osc}}(\rho)=\frac{\mu^2\omega^2R^2}{\hbar^2} \tanh^2\rho&=&  \kappa^4 R^2 \left(1-\frac{1}{\cosh^{2}\rho}\right),\quad
\frac{\mu^2\omega^2}{\hbar^2}=\kappa^4 .
\label{defbeta}
\end{eqnarray}
In so doing amounts to
\begin{eqnarray}
\left[-
 \frac{1}{R^2}{\mathcal C}(\rho,\varphi) + \kappa^4R^2 \tanh^2\rho\right]
\Psi_n^{a\kappa} (\rho, \varphi)&=&\left({\epsilon}_{n}^{a \lambda }\right)^{Higgs} \Psi_n ^{a\kappa } (\rho ,\varphi),\nonumber\\
\left({\epsilon}_{n}^{a \lambda }\right)^{Higgs} &=&\frac{2\mu}{\hbar^2}\left(E^{a\kappa}_n\right)^{{ Higgs}}.
\label{ET1PT}
\end{eqnarray}

Changing variable to $\Psi^{a\kappa}_n(\rho,\varphi)==\psi_n^{a\lambda}(\rho)e^{im\varphi}/\sqrt{\sinh\rho}$,
with $\lambda$ to be defined shortly below, allows again to cast 
 the Higgs oscillator potential problem on {\bf H}$^2_{+R}$ as the following one-dimensional Schr\"odinger equation,
which,  in units of  $\hbar^2/(2\mu)=1$, i.e. in units of $\lbrack R^2\rbrack $=fm$^{-2}$ reads,
\begin{eqnarray}
\frac{1}{R^2}\left[ -\frac{{\mathrm d}^2}{{\mathrm d}\rho^2} +  {\mathcal V}_{PTII}(\rho)\right]\psi_n^{a\lambda } (\rho)=
\left({\epsilon}_{n}^{a \lambda }\right)^{Higgs}\psi_n^{a \lambda } (\rho),&&\label{HOADSPT}\\
{\mathcal V}_{PTII}(\rho)  =\frac{a(a-1) }{\sinh^2\rho }-\frac{\lambda(\lambda +1)}{\cosh^2\rho}
+\kappa^4 R^2 +\frac{1}{4R^2},&&\label{PTIIpot}\\
a=\pm |m| +\frac{1}{2}, \quad \lambda(\lambda+1)=\kappa^4R^4, \quad \lambda =-\frac{1}{2} \pm |s|, \,\, |s|= \sqrt{\kappa^4R^4 +\frac{1}{4}}.&&
\label{HOADS}
\end{eqnarray}
The potential ${\mathcal V}_{PTII}(\rho)$ (modulo the two additive constants) is known \cite{CooperKhare}, \cite{Wipf}, \cite{Suprami}  
under the names of either  the ``generalized'', or the ``second''  hyperbolic  P\"oschl-Teller potential, abbreviated, 
P\"oschl-Teller II .
In other words, in the hyperbolic variable, the Higgs oscillator on {\bf H}$_{+R}^2$ amounts equivalent to the second P\"oschl-Teller potential on a hyperbolic geodesic.
Notice, that such a potential  can be obtained within the  AdS/CFT framework  as a Wilson loop potential with a 
Rindler universe as a background \cite{Rindler}. 
The related energies, $\left({\epsilon}_{n} ^{a\lambda} \right)^{{ PTII}}$, are given by \cite{Wipf}, \cite{Suprami}
\begin{eqnarray}
\left({\epsilon}_{n} ^{a\lambda} \right)^{{PTII}}=-\frac{1}{ R^2}\left(\lambda -a- 2n \right)^2 
=-\frac{1}{R^2}{\Big(}\sqrt{\kappa^4R^4+\frac{1}{4}} &-&{\Big(}\frac{1}{2}+a+2n{\Big)}{\Big)}^2,\nonumber\\
n=0,1,2...<\frac{\lambda -a }{2}\,\,&=&\frac{|s|-|m|-1}{2},
\label{step1}\\
\left({\epsilon}_{n} ^{a\lambda} \right)^{{ Higgs}}=\left({\epsilon}_{n} ^{a\lambda} \right)^{{PTII}}&+&\kappa^4R^2 +\frac{1}{4 R^2}.
\label{spectrum_secondPT}
\end{eqnarray} 
Correspondingly, the wave functions are
\begin{equation}
\psi_n^{a\lambda}(\rho)\sim  \sinh^{1-a}\rho \, \cosh^{\lambda +1} \rho \,\, _1 F_1 \left(-n, \lambda -a -n+1, \frac{3}{2}-a,-\sinh^2\rho \right).
\label{WAFUPT_gen}
\end{equation}
The spectrum (\ref{step1}) no longer shares the $so(1,2)$ Lorentz symmetry with the unperturbed excitations in
(\ref{Eckrt_enrgs}), (\ref{ET1}).

\section{ Light-Front Holography  as the small hyperbolic angle  limit of  the 
Higgs oscillator on $H_{+R}^2$.  Conformally symmetric quark-diquark models
in the ultraviolet and infrared regimes of QCD}

\subsection{Reproducing the conformal holographic interaction}

\begin{figure}
\resizebox{0.80\textwidth}{7.5cm}
{\includegraphics{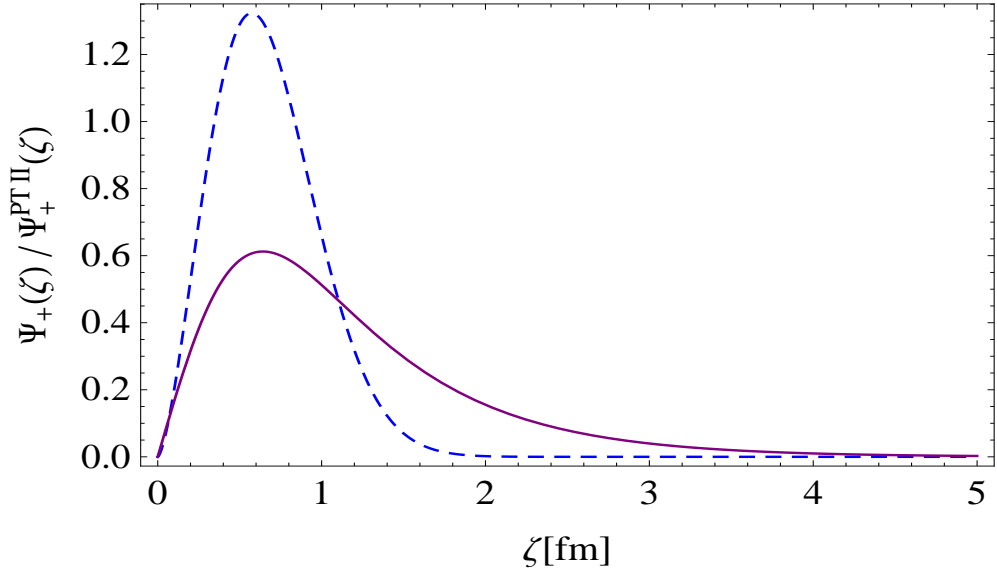}}
\caption{ Comparison of the  hyperbolic ground state wave function  $\psi^{PTII}_+(\zeta)=\psi_{01\left(\frac{5}{2}\right)}(\zeta)$ in (\ref{gst_SecondPT}) (solid line) to  the 
LFH wave function of equal quantum numbers, $\Psi_+^{01}(\zeta)$, in (\ref{gst_test}) (dashed line).
We suppressed the upper indexes in the LFH wave function
for the sake of  keeping same notations as in \cite{GdTSBJ} .
The $\kappa$ and $\nu$ parameters have been taken same  in both functions,   
and carry the values  $\nu=1$, and $\kappa$= 2.14 fm$^{-1}$, respectively. The value of the radius of the hyperbolic
function has to be consistent with the condition in (\ref{step1}), i.e. with $|s|>2n+|m|+1$, 
which determines the number of bound states within the potential under discussion. 
The radius is then  deduced from our adopted
$|s|=\frac{5}{2}$ value  as $R=0.728$ fm.
 These are the parameters to be used in all subsequent calculations. 
\label{figure1}
}
\end{figure}
In the limit of small hyperbolic angles, which formally coincides with the limit of  an increasing radius, $R\to \infty $,  the equations
(\ref{HOADSPT})- (\ref{PTIIpot}) approach, modulo additive constants, the 
Light-Front Holographic equation (\ref{Gl1}).
Indeed,  it is not difficult to see that the  $\csch^2\rho$ term approaches 
the inverse square distance potential, while the $\sech^2\rho$ potential goes to the harmonic oscillator according to,
\begin{eqnarray}
\frac{1}{ R^2}\frac{a(a-1)}{\sinh^2\rho} &\longrightarrow & a(a-1)\frac{1}{ \zeta^2},\quad \rho=\frac{\stackrel{\frown}{z}}{R},\quad
\stackrel{\frown}{z}\, \stackrel{R\to\infty}{\longrightarrow}\,  \zeta,
\label{lim1}     \\
\kappa^4R^2{\tanh^2\rho} &\longrightarrow &   \kappa^4R^2\frac{\stackrel{\frown}{z}^2}{R^2}\longrightarrow \kappa^4 \zeta^2. 
\label{lim2}
\end{eqnarray}
Here, use has been made of the fact that the arc of the hyperbolic geodesic, $\stackrel{\frown}{z}$, approaches the segment, $\zeta$,  
of a straight line. In effect, the form of the conformal interaction in (\ref{Gl1}), (\ref{Gl1_n}) is recovered as,
\begin{equation}
R\to \infty:\quad \frac{1}{R^2}\frac{a(a-1) }{\sinh^2\rho }
-\frac{1}{ R^2}\frac{\kappa^4R^4}{\cosh^2\rho}+\kappa^4 R^2   \longrightarrow \frac{a(a-1)}{\zeta^2} +\kappa^4 \zeta^2.
\end{equation}
This is in accord with (\ref{Gl1}), for  $a$  becoming  $\left(\nu +\frac{1}{2}\right)$.
With that, the equivalence of the differential equations (\ref{HOADSPT})--(\ref{HOADS}) and (\ref{Gl1}) (modulo the additive constants) 
 has been demonstrated.

\subsection{Reproducing the  LFH   energies}
Upon substituting again in (\ref{step1})-(\ref{spectrum_secondPT}) the $\lambda $ value from (\ref{HOADS}), 
the contraction limit of the energy is  easily worked out as (see \cite{Alonso} for a related consideration)
\begin{eqnarray}
\lim_{R\to \infty}\left( \epsilon_{n}^{a\lambda}\right)^{Higgs}= 
\lim _{R\to \infty}\left[ 
-\frac{1}{R^2}
\left( \sqrt{\kappa^4R^4 +\frac{1}{4}}-\left(\frac{1}{2}+ a +2n\right) \right)^2 
+\kappa^4R^2 
+\frac{1}{R^2}\right]&&\nonumber\\
\to
\lim_{R\to \infty}
\left[ \kappa^4R^2 
-\frac{\left( 2n+a
 +\frac{1}{2}\right)}{R^2}\left( \left(2n+a+\frac{1}{2}\right) -2
\sqrt{{\kappa^4R^4} +\frac{1}{4}} +{\kappa^4R^4} +\frac{1}{4}
\right)\right]&&\nonumber\\
\longrightarrow  2\kappa^2\left(2n+a +\frac{1}{2} \right)
=4\kappa^2\left(n+\frac{|m|+1}{2}\right),&&
\label{bo2}
\end{eqnarray}
with $a$ being substituted by $a=|m|+\frac{1}{2}$.
We now recall  that the energy in (\ref{bo2}) has the dimensionality of fm$^{-2}$, i.e. 
$\left[ \left({\epsilon}_{n}^{a \lambda }\right)^{Higgs}\right]={{\mathrm f}{\mathrm m}}^{-2}$,
and coincides with the dimensionality of the quadratic energy in (\ref{SUSY_sols}).
Upon identifying  $|m|$ with $\nu$, taking into account the rational units, 
$\hbar=1=2\mu$, and finally incorporating the additive constant $c_+^\nu=2\kappa^2(\nu +1)$ from (\ref{Gl1}) 
into (\ref{bo2}), one finds that  the contraction limit of the second P\"oschl-Teller II potential on the 
hyperbola {\bf H}$_{+R}^1$ generates same spectrum as the Light-Front framework in (\ref{Gl1}), (\ref{Gl1_n}), namely,

\begin{eqnarray}
\lim_{R\to \infty}\left( \epsilon_{n}^{a\lambda}\right)^{Higgs}+2\kappa^2(|m|+1)&=&
 4\kappa^2\left(n+{|m|+1}\right)={E}^2,\, |m|=\nu.
\label{Higgs_LFH}
\end{eqnarray}
In conclusion, we demonstrated that the Light-Front holographic
equation and its spectrum can, alternatively to AdS$_2$/CFT$_1$,  be also traced back to the decreasing curvature
(equivalently, small hyperbolic angle) limit of  
the Higgs oscillator on the hyperbolic plane {\bf H}$_{+R}^2$ ,  
reduced to the one-dimensional P\"oschl-Teller potential on the hyperbola 
${\mathbf H}_{+R}^1$. In this fashion, the one dimensional Schr\"odinger equation of the holographic Light-Front QCD
in the ${\mathcal R}^{1}$ space of the $\zeta$ variable has been recovered.

\subsection{Reproducing the LFH wave functions}

In the following, the potential parameters $a$ and $\lambda $ in (\ref{HOADS}), which have been defined in (\ref{PTIIpot}) 
via their squares according to,   $a(a-1)=|m|^2-\frac{1}{4}$, now with $a=\pm |m|+\frac{1}{2}$, and 
$\lambda(\lambda +1)=\kappa^4R^4$, with $\lambda =-\frac{1}{2}\pm \sqrt{\kappa^4R^4 +\frac{1}{4}}$
will be chosen as,
\begin{equation}
a=-|m| +\frac{1}{2}, \quad \lambda =-\frac{1}{2}- |s|, \quad |s|=\sqrt{{\kappa^4R^4}+\frac{1}{4}}.
\label{PTII_prms}
\end{equation}
This choice is compatible with  the sign inversions of all the quantum numbers entering the definition of the energy, 
$\left( \epsilon _n^{a\lambda}\right)^{PTII}$, in (\ref{step1}).
We now switch to  more explicit (with respect to  ~(\ref{HOADS})) labellings of the wave functions and energies as,
\begin{equation}
\psi_n^{a\lambda }(\rho) \longrightarrow \psi_{n|m||s|}(\rho), \quad 
\epsilon_{n}^{a\lambda}\longrightarrow \epsilon_{n|m||s|}.
\end{equation}
In accord with (\ref{WAFUPT_gen}) these  wave functions now read,
\begin{equation}
\psi_{n|m||s|}(\rho) \sim \cosh^{\lambda +1}\rho \, \sinh^ {|m|+\frac{1}{2}}
_2F_1\left(-n,\left(-|s|-n+|m|+1\right),|m|+1,-\sinh^2\rho\right).
\label{wafu_secondPT}
\end{equation}
We first focus on the ground state wave function which is especially simple,
\begin{equation}
\psi^{PTII}_+\left({\zeta }\right): \Rightarrow \,\,
\psi_{0|m||s|}(\rho)\sim \cosh^{-|s|+\frac{1}{2}}\rho \,\sinh^{|m|+\frac{1}{2}}\rho, \quad \rho=\frac{\zeta}{R}.
\label{gst_SecondPT}
\end{equation}
In making use of the approximation \cite{Alonso},
\begin{equation}
\cosh \rho \approx \exp  \left( \tanh^2\frac{\rho}{2}\right)
\approx 1+\frac{1}{2}\rho^2\approx \exp\left(\frac{\rho^2}{2} \right),
\label{hypcos_exp}
\end{equation}
 the $R\to \infty$ limit of the $\cosh^{-\lambda +\frac{1}{2}}\rho$ factor easily calculates as,
\begin{eqnarray}
\lim _{R\to \infty}\cosh^{-|s|}\rho &\longrightarrow& \exp \left( -|s|\tanh^2\frac{\rho}{2}\right)\longrightarrow 
\exp \left(-|s|\frac{\rho^2}{2} \right)\nonumber\\
&\longrightarrow& \exp\left( - \kappa^2R^2\frac{\stackrel{\frown}{z}\, ^2}{2R^2}\right)
\longrightarrow \exp \left(-\frac{\kappa^2\zeta^2}{2} \right),
\end{eqnarray}
where once again the arc of the geodesic,  $\stackrel{\frown}{z}$, approaches the segment, $\zeta$,  of a straight line in the $R\to \infty$ limit. The $\left(\sinh^2\rho\right)^{\frac{m}{2} +\frac{1}{4}}$ factor behaves as,
\begin{equation}
\lim_{R\to \infty}\left(\sinh^2\rho\right)^{\frac{|m|}{2} +\frac{1}{4}}\sim  \left(\zeta^2\right) ^{\frac{|m|}{2}+\frac{1}{4}}.
\label{sinchfac}
\end{equation}
Putting all together, the diminishing curvature limit of  the ground state wave function $(|m|=1)$ of the second P\"oschl-Teller potential is 
found as expected, 
\begin{equation}
\Psi_+^{01}(\zeta):\Rightarrow  \,\, \lim_{|s|\to \infty}\psi_{0|1||s|}(\rho)=N
e^{-\frac{\kappa^2\zeta^2}{2}}\left(\kappa^2\zeta^2\right)^{\frac{1}{2}+\frac{1}{4}}, \quad |s|\approx \kappa^2 R^2,
\label{gst_test}
\end{equation}
where $N$ is a  normalization constant.
In this way, the $\Psi_+^{01}(\zeta)$  ground state in  the Light-Front Holography following from (\ref{Gl2}) for $n=0$, and $\nu=1$
is recovered upon identifying as usual $|m|$ with $\nu$ (see Fig.~\ref{figure1}). It is this wave function that is relevant for the description of the proton electric charge form factor below.
Similarly, the remaining states can be analyzed. Indeed, in using the relationship between $_2F_1$ and $_1F_1$ hypergeometric functions,
and the relationship of the Laguerre polynomials to $_1F_1$ \cite{Abramovic}, \cite{jdcook},
one finds the polynomial part of the wave function in (\ref{wafu_secondPT}) expressed in the following way in 
the $|s|\to \infty$  limit: 
\begin{eqnarray}
_2F_1(-n,-|s|+n+1,|m|+1,-\sinh^2\rho)\stackrel{|s|\to \infty}{\longrightarrow} \, _1F_1\left(-n,|m|+1,|s|\sinh^2\rho\right)&&\nonumber\\
\stackrel{|s|\to \infty}{\longrightarrow}\,  _1F_1(-n, |m|+1,\kappa^2\zeta^2)\sim L^{|m|}_n(\kappa^2\zeta^2).&&
\label{bong}
\end{eqnarray}
 Comparison to (\ref{Gl2}), and upon identifying once again $|m|$ with $\nu$,   confirms that also the polynomial part for the ``curved'' wave functions in (\ref{wafu_secondPT})
approach the correct Laguerre polynomials defining the wave functions of the Light-Front Holographic equation, as it should be.
In this fashion, the exact solutions to (\ref{Gl1}) from (\ref{Gl2}) are reproduced in the contraction limit
of {\bf H}$_{+R}^1$ to a straight line on a cone (see also \cite{Alonso} where the contraction limit of the Higgs oscillator on the hyperbola has been elaborated). 
As long as the  inverse-square distance plus harmonic oscillator potential in one dimension represents a conformally invariant interaction, 
because the associated  Hamiltonian is realized in terms of  the  $so(2,1)$ algebra elements in (\ref{conf_alg}),
a conformally symmetric  interaction is found  in the contraction limit.

\subsection{Reproducing the LFH  proton electric charge form factor from the Fourier-Helgason hyperbolic wave transform}
\begin{figure}
\resizebox{0.80\textwidth}{7.5cm}
{\includegraphics{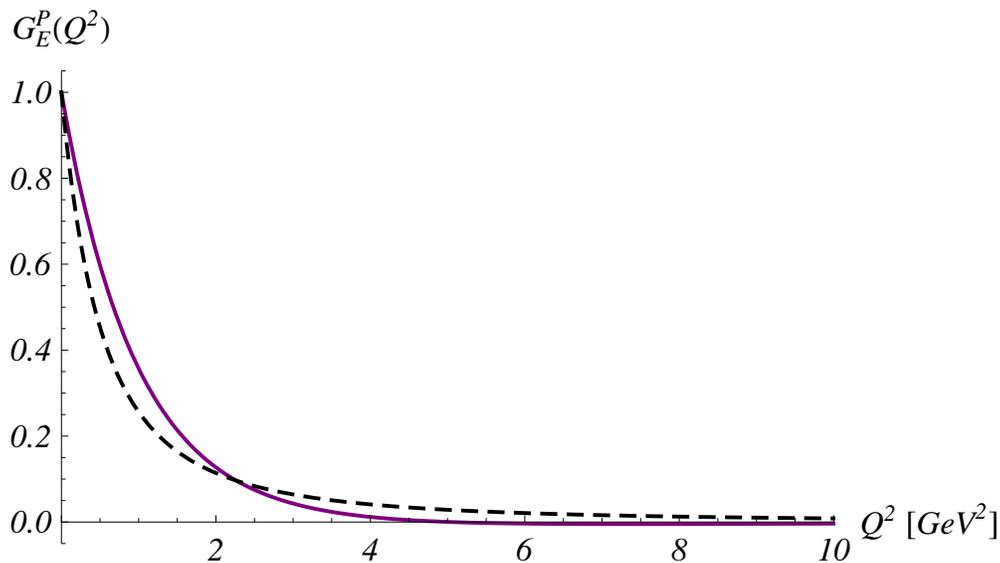}}
\caption{ Side by side comparison of the  proton electric charge form factor, $G^p_E(Q^2)$, predicted by two distinct conformal quark-diquark
models, the one elaborated here and suited for the ultraviolet regime of QCD (solid line), and another one, earlier presented in \cite{TQC} and relevant in the 
infrared regime (dashed line). 
The calculation in the ultraviolet has been performed  in (\ref{FnRslt}) in terms of the small angle approximation to the 
ground state wave function of the  P\"oschl-Teller II potential and using a Fourier-Helgason transform of the charge-density, while the one in the infrared is based  on the wave function of the trigonometric Rosen-Morse potential, a $\csc^2+\cot$ interaction and the equation (\ref{RMt_FF}) given in the Appendix. 
 \label{figure2}
}
\end{figure}

\begin{figure}
\resizebox{0.80\textwidth}{7.5cm}
{\includegraphics{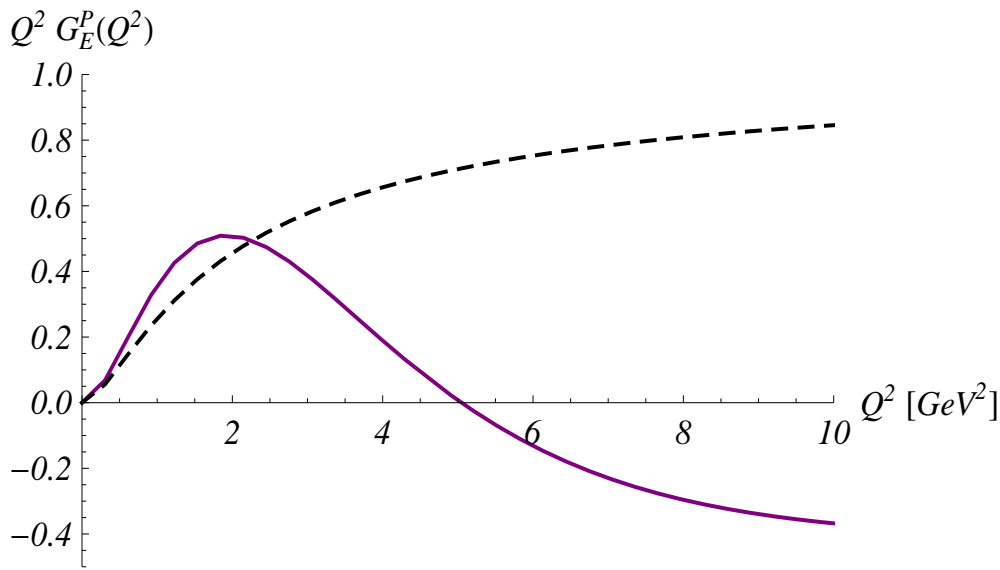}}
\caption{ Comparison of the behaviors of $Q^4 G^p_E(Q^2)$, calculated in the ultraviolet with the
small angle approximation to ground state wave function of the second hyperbolic  P\"oschl-Teller potential in
(\ref{gst_SecondPT}) (solid line),  and in the infrared with the one corresponding to the conformal trigonometric Rosen-Morse potential in \cite{TQC} (dashed line). Notice the correct increase near origin,
shared by both form factors.   
\label{figure3}
}
\end{figure}

\begin{figure}
\resizebox{0.80\textwidth}{7.5cm}
{\includegraphics{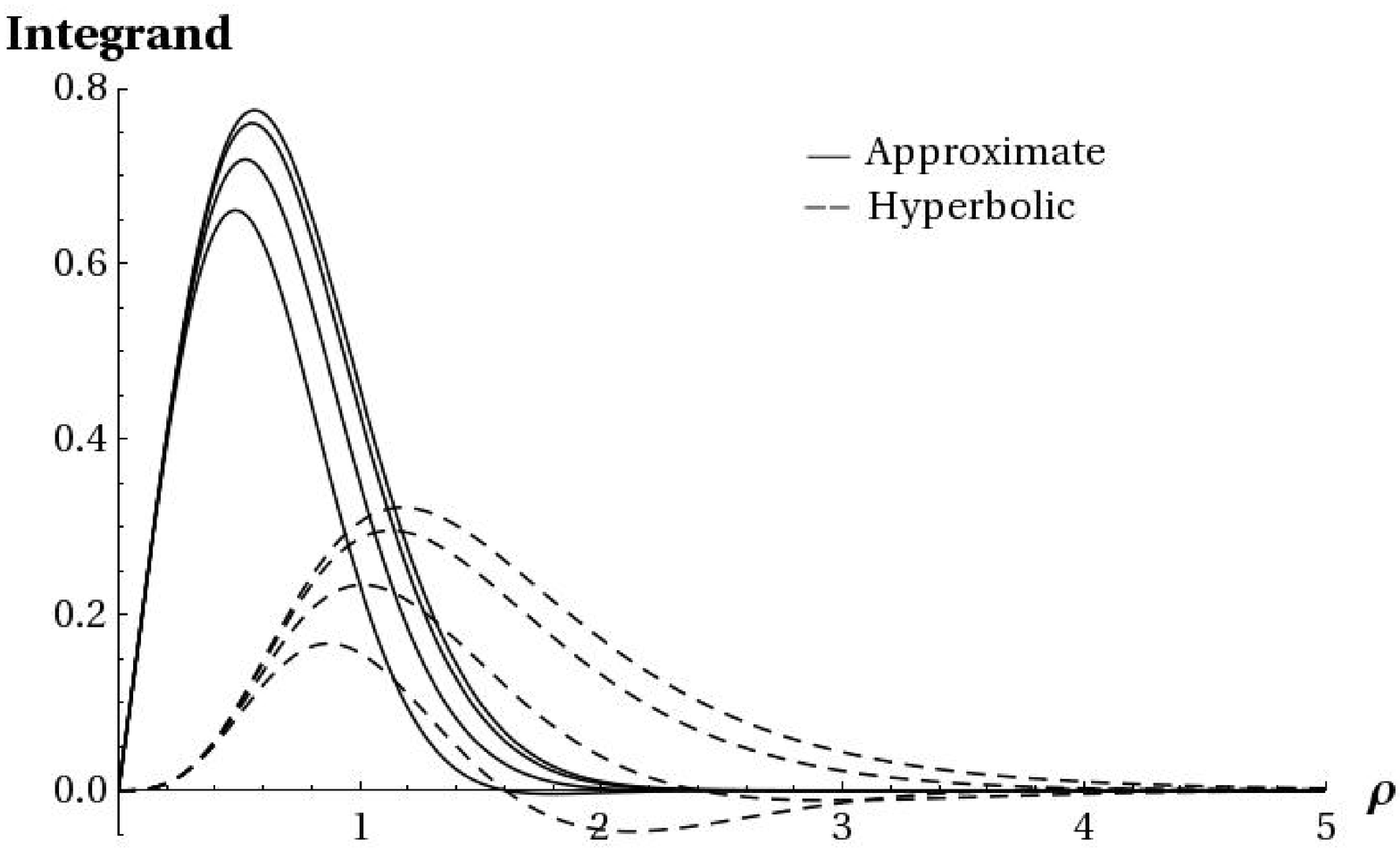}}
\caption{ Approximate (\ref{FoF1_1}) (solid line) and exact  (\ref{hyperb_krn})-(\ref{FoF1}) (dashed line)  integrands.
The various integrands correspond from bottom to top to decreasing  $Q=3,2,1,0$ values (in GeV). 
The small differences between the areas under the  solid--, and dashed line curves of same $Q$'s are illustrative of the reasonable accuracy of the small angle approximation in the evaluation of the Fourier-Helgason integral transform.
 \label{figure4}
}
\end{figure}

The proton electric charge form factor in the Conformal  Light-Front Holographic QCD is well known and its calculation will not be reproduced here.
Instead we explore utility of the hyperbolic wave function in (\ref{gst_SecondPT}) for the description of the same observable.
The normalized ground state proton wave function reads,
\begin{equation}
\psi_{0|m||s|}(\rho)=\sqrt{\frac{2\Gamma (|s|)}{\Gamma(|m|+1)\Gamma(|s|-|m|-1)}}\cosh^{-|s|+\frac{1}{2}}\rho\sinh^{|m|}\rho,
\label{protongst}
\end{equation}
Our calculation refers to a quark-diquark system of reduced mass $\mu$, as commented immediately after eq.~(\ref{Gl1}) above.
The adequate momentum space, dual to the curved position space on the hyperboloid, is obtained via the so called Fourier-Helgason
integral transform \cite{IvaBogd} based on the hyperbolic waves, also referred to in \cite{Alonso} as Shapiro functions,
$\Phi_{\mathbf p}^{(D)}({\mathbf r})$, with $D$ standing for the dimensionality of the surface.
Specifically for the hyperboloid under investigation, $D=2$, they read,
\begin{eqnarray}
\Phi_{\mathbf p}^{(2)}({\mathbf r})&=&\left( \frac{x_0}{R} -
\frac{\hat{ \mathbf p}\cdot {\mathbf r}}{R}\right) ^{-\frac{1}{2}-ipR}, \quad 
{\mathbf r}=\left(\begin{array}{c}
x_1\\
x_2
\end{array}
\right)=R{\hat {\mathbf r}}\sinh \rho.
\label{Shp1}
\end{eqnarray}    
Here, ${\hat {\mathbf p}}$ is the unit vector along the direction of the momentum ${\mathbf p}$ and  tangent to the surface under consideration, $p$ is its magnitude, 
while ${\mathbf r}$ is the two-dimensional radius vector in the Euclidean $(x_1,x_2)\in {\mathcal R}^2$ plane. In making use of (\ref{ET2}), the latter equation becomes,
\begin{eqnarray}
\Phi_{\mathbf p}^{(2)}({\mathbf r})&=&\left( \cosh\rho -\hat{ \mathbf p}\cdot \hat{\mathbf r}\sinh\rho \right) ^{-\frac{1}{2}-ipR}, \nonumber\\
&=&\left( e^{-\rho\hat{ \mathbf p}\cdot \hat{\mathbf r}}\right)^{-\frac{1}{2}-ipR}, \quad
=e^{\frac{\rho}{2}\hat{ \mathbf p}\cdot \hat{\mathbf r}}e^{i{\rho}{ \mathbf p}\cdot R\hat{\mathbf r}}.
\label{Shp2}
\end{eqnarray}    
In recalling that the hyperbolic angle $\rho$ in (\ref{Shp2}) expresses in terms of the arc along a geodesic and the constant radius as,
$\rho =\frac{\stackrel{\frown}{z}}{R}$,  amounts to the following equivalent expression for the Shapiro function (also see Ref.~\cite{IvaBogd} for more details),
\begin{eqnarray}
\displaystyle{
\Phi_{\mathbf p}^{(2)}({\mathbf r})=e^{\frac{\stackrel{\frown} {z}
}{2R}\hat{ \mathbf p}\cdot \hat{\mathbf r}}e^{i\stackrel{\frown}{z}{ \mathbf p}\cdot \hat{\mathbf r}}, \quad {\hat {\mathbf p}}\cdot \hat {{\mathbf r}}=\cos\varphi,}
\label{Shp3}
\end{eqnarray}
where $\varphi$ is taken as the azimuthal angle in {\bf H}$_{+R}^2$ in (\ref{ET2}).
The latter expression shows that in the $R\to \infty$ limit, in which the hyperbolic geodesic stretches to a line
of a direction given by ${\hat {\mathbf r}}$ in ${\mathcal R}^2$, 
the arc $\stackrel{\frown}{z}$ approaches a corresponding line segment, $\zeta$.  In this limit,           
the hyperbolic plane waves (the Shapiro functions) approach ordinary plane waves in a two-dimensional flat space 
and the integral transform correspondingly evolves to an ordinary two-dimensional  Fourier transform, 

\begin{equation}
\lim _{R\to \infty}\Phi_{\mathbf p}^{(2)}({\mathbf r})\longrightarrow e^{i{ \mathbf p}\cdot {\mathbf r}}.
\label{Shp4}
\end{equation}
The Shapiro functions are normalized according to \cite{Alonso} as,
\begin{eqnarray}
\frac{R}{2\pi}\int _{{\mathbf r}\in {\mathcal R}^2}
\frac{{\mathrm d}^2{\mathbf r}}{x_0}
\Phi ^{(2)}_{{\mathbf p}}\, ^\ast ({\mathbf r}) \Phi_{{\mathbf p}^\prime}^{(2)}(\mathbf r)&=&
N^{(2)}(p)\delta^{(2)}({\mathbf p}-{\mathbf p}^\prime),\nonumber\\
\frac{R}{2\pi}\int _{{\mathbf p}\in {\mathcal R}^2}
\frac{{\mathrm d}^2{\mathbf p}}{N^{(2)}(p)}
\Phi ^{(2)}_{{\mathbf p}}\, ^\ast ({\mathbf r}) \Phi_{{\mathbf p}}^{(2)}({\mathbf r}^\prime )&=&
\delta^{(2)}({\mathbf r}-{\mathbf r}^\prime),\nonumber\\
\label{Shapiro}
\end{eqnarray}
In terms of the Shapiro functions, the momentum space for the hyperboloid is defined according to \cite{Alonso}, \cite{IvaBogd},
\begin{eqnarray}
F({\mathbf p})&=&\frac{R}{2\pi}\int _{{\mathbf r}\in {\mathcal R}^2}
\frac{{\mathrm d}^2{\mathbf r}}{x_0}
\Phi ^{(2)}_{{\mathbf p}}\, ^\ast ({\mathbf r}) f(\mathbf r),\nonumber\\
f({\mathbf r})&=&\frac{R}{2\pi}\int _{{\mathbf p}\in {\mathcal R}^2}
\frac{{\mathrm d}^2{\mathbf p}}{N^{(2)}(p)}
\Phi _{{\mathbf p}}^{(2)} ({\mathbf p}) F({\mathbf p}).
\label{Shapiro_transforms}
\end{eqnarray}
Within this framework, the proton electric form factor on the hyperbolic space is obtained  as a Fourier-Helgason 
transform of the  squared ground state wave function in (\ref{protongst}) (with the ``curved'' wave  function containing a
$\sqrt{\sinh\rho}$ less than the Schr\"odinger one according to (\ref{ET2}),  as 
\begin{eqnarray}
G_E^p({Q^2})=
R^2\int_{0}^{\infty} {\mathrm d}\rho\,\frac{\sinh\rho}{\cosh\rho} 
\frac{|\psi_{0 1|s|}(\rho)|^2}{\sinh\rho}  \int_{-\pi}^{+\pi} {\mathrm d}\varphi
e^{\frac{\rho\cos\varphi}{2}}e^{iQR\rho\cos\varphi}&&\label{hyperb_krn}\\
\approx\int_{0}^{\infty}
{\mathrm d}\, \rho 
\cosh^{-2|s|+1}\rho \,\sinh^2\rho \,
e^{\frac{\rho}{2}}\int _{-\pi}^{+\pi}
{\mathrm d}\varphi 
e^{iQR\rho\cos\varphi}, 
\,\, \cos\varphi={\hat {\mathbf q}  }\cdot {\hat {\mathbf r}}, &&
\label{FoF1}
\end{eqnarray}
where ${\hat {\mathbf q}  }$ is the unit vector along the transferred momentum, and $Q^2=-{\mathbf q}^2$.
In order to obtain a closed value for $G_E^p({Q^2})$, we ignored presence of  $|\cos\varphi| \leq 1 $ in the real exponential factor,
i.e, we  replaced $e^{\frac{\rho}{2}\cos\varphi}$ by $e^{\frac{\rho}{2}}$. 
This allows us to carry out the integration over the azimuthal angle as \cite{AW} 
\begin{equation}
\frac{1}{2\pi}
\int_0^{2\pi}e^{i\rho QR\cos\varphi}{\mathrm d}\varphi =J_0(Q R\rho), \quad Q=|{\mathbf q}|.
\end{equation}
Next we are interested in relatively small angles in view of the
rapid decrease of the ground sate wave function with $\rho$ as illustrated by Fig.~\ref{figure1}.
With that in mind, we keep only the lowest terms in the $\exp (\rho/2)$ expansion, and approximate the
$\sinh\rho$ from the integration volume by $\rho$, which allows us to convert  
the Fourier-Helgason transform to a Hankel transform.
{}Finally, in systematically making use of eq.~(\ref{hypcos_exp}) to replace everywhere $\cosh\rho$ by $\exp (\rho^2/2)$,
and make use of $\tanh \rho \stackrel{R \to \infty}{\longrightarrow}1$
 allows to simplify the equation (\ref{FoF1}) as,
 \begin{eqnarray}
G_E^p({Q^2})&=&
R^2\int_{0}^{\infty} {\mathrm d}\rho\,e^{-\frac{3}{2}\rho^2}\left( 1+\frac{\rho}{2}\right)
\rho \, J_0(QR\rho).
\label{FoF1_1}
\end{eqnarray}
The latter integral takes in closed form by means of the symbolic software Mathematica and reads,

\begin{equation}
G_E^p({Q^2})=\frac{R^2}{2}\, e^{-\frac{Q^2R^2}{6}}
\left(1-\frac{1}{12}\sqrt{\frac{\pi}{6}}\left[ (Q^2R^2-6)I_0\left(\frac{Q^2R^2}{12} \right)-Q^2R^2I_1\left( \frac{Q^2R^2}{12}\right)\right] \right),
\label{FnRslt}
\end{equation}
where $I_0$ and $I_1$ denote the modified Bessel functions of zeroth and first degree, correspondingly.
\begin{itemize}
\item In this fashion, the proton electric charge form-factor has been calculated within a hyperbolic quark-diquark  
model suited for dynamics in the  ultraviolet and respecting in the flat-space limit the conformal symmetry. 

\end{itemize}
The observable obtained is displayed in Fig.~\ref{figure2}, where it has been compared to same physical entity, earlier calculated in the infrared
in \cite{TQC} once again employing  a conformal  quark-diquark model, namely the one  based on the trigonometric Rosen-Morse potential.
The  Figure~ \ref{figure3}  shows the behavior of both form factors under discussion with the increase of the momentum transferred.
{}Finally, in Fig.~\ref{figure4} we display the approximate and exact hyperbolic integrands corresponding to  (\ref{FoF1}).
The model employed in the ultraviolet and elaborated here is placed on a non-compact (1+2) Minkowski  space, ${\mathbf H}_{+R}^2$
and in accord with the relativistic nature of the ultraviolet (ultra-relativistic) regime of QCD.
The small hyperbolic approximation used guarantees conformal symmetry in accord with LFH. Instead, 
the previously developed model in the infrared, that has been used as a comparison, is placed on the compact version of 
${\mathbf H}_{+R}^3$ which is the  hyperspherical  space, $S^3_R$, as more appropriate in the near rest-frame regime.
The conformal symmetry of the $\csc^2$ piece is obvious from the fact that it appears as a part of the Laplace-Beltrami operator on $S^3_R$ and defines the  
free motion there.  The  perturbation by the  cotangent term, a harmonic function on the surface under consideration, also known as ``curved'' 
Coulomb perturbance, conserves the conformal algebra. The reason is that the Hamiltonian describing such a motion can be cast in the form of a  
Casimir invariant of a non-trivial representation of the $so(4)$ sub-algebra of the conformal algebra $so(2,4)$, a result due to  \cite{PalKi}.  
Both form-factors obviously compare in  quality and  go practically through the data, not shown here (see \cite{Data_FF},\cite{TQC} and references therein).
The coincidence of the two form-factors strongly points on conformal symmetry as the chief designer of  quark dynamics in the two extreme QCD regimes.

{}Finally, the behavior  of $Q^4 G^p_E(Q^2)$ near origin displayed in Fig.~\ref{figure3}  is once again illustrative of the proximity  
of the small hyperbolic angle approximation to the  P\"oschl-Teller II wave functions to the conformally invariant Light-Front Holographic ones. 
At high momentum transfers, however, only the form factor corresponding to the infrared shows the 
desired scaling property. The divergence of the  form-factor in our hyperbolic model is understandable in view of the circumstance that
its conformal symmetry is limited to small angles.   

\section{Conclusions}

In this work we studied the Higgs oscillator on the hyperbolic plane, reduced to the 1D Schr\"odinger equation
describing motion on a hyperbola perturbed by the second hyperbolic P\"oschl-Teller potential.
In considering the decreasing  curvature limit, equivalent to small hyperbolic angles, 
we found in eqs.~(\ref{lim1})--(\ref{lim2}) that this  potential amounted to the one dimensional 
conformally  symmetric interaction consisting of an inverse square distance plus a square distance potentials.
In this fashion, the Lorentz symmetry of the initial hyperbolic interaction (i.e. $ \ell (\ell +1)/\sinh^2\rho$ in 
(\ref{ET1})), that has been broken by the perturbation due to the Higgs oscillator in (\ref{HOADSPT})-(\ref{HOADS}),  
 to the end has been converted to conformal interaction  in the flat space  limit.
Comparison of the Light Front Holographic  ground state wave function in (\ref{gst_test}) to the hyperbolic  one 
in (\ref{protongst}) was presented in Fig.~1.
We observed that for a relatively short radius, of the order of $1$ fm, the wave functions coincided quite reasonably.
Such is due to the fact that the most essential part of the LFH wave function is  located in the small distance region, comparable 
to small angles on the curved surface. 
 Within same scheme, we also studied the proton electric charge form factor, shown in Figs.~2 and ~3, 
and found again a pretty realistic data description in terms of a Fourier-Helgason transform of the charge density.
In result, we found that essential outcomes of the Light-Front Holographic QCD in the ultraviolet can independently be 
backed up by a hyperbolic relativistic  quark-diquark ($q-(qq$)) model, consistent with conformal symmetry.
The model  parallels in the ultraviolet another one, earlier developed in the infrared, and in which
the $q-(qq)$ system, placed on the compact space of the hypersphere $S^3_R$,  interacts via the
``curved'' Coulomb potential, a cotangent function of the second polar angle. This function is part of  
the standard trigonometric Rosen-Morse potential \cite{Manning_Rosen} (see the Appendix for details). 
The cotangent potential has been shown in \cite{PalKi} to be generated by the Casimir invariant of the $so(4)$ algebra in a representation 
nonequivalent to the canonical, a reason for which it represents a conformal interaction.
In this fashion, we showed that  the conformal symmetry,  
the decisive mechanism in shaping the light flavor hadron spectra  and the  electric charge form factor,     
can be treated on similar terms in the two extreme regimes of QCD. 
Spaces of constant curvatures, are known to efficiently absorb  part of the 
interactions into the free  motion,  and frequently happen to provide convenient scenarios for the description of complex dynamical systems \cite{IvaBogd}.
Such models can further be upgraded by perturbations of the free motions and in this fashion  the 
level of phenomenological description of coupled systems can be improved.  
The inverse radius of the {\bf H}$^2_{+R}$ space of constant curvature considered here provides a second scale in addition to the strength of the perturbing potential and can  be employed  in the definition of chemical potential along the lines of Refs.~\cite{Ebert},\cite{Anim}.

\begin{itemize}
\item In summary,  in the ultraviolet, as well as in the infrared,  the conformal symmetry of the Schr\"odinger equations 
has been realized in the presence of  two external length scales, brought about by the respective potential strengths,  
on the one side, and by the curvatures of 
of the respective relevant hosting curved mannifolds, on the other side.
\end{itemize}
The findings reveal  the possibility  that a new generation of quark models based on trigonometric or hyperbolic SUSY-QM potentials  is capable of 
capturing  notably more of the essential field-theoretical--, and first-principle AdS/CFT aspects  of QCD in the infrared and the ultraviolet than 
models  based on the traditional power potentials. 

\section{Appendix: The trigonometric Rosen-Morse potential}
In this Appendix we summarize for the sake of completeness of the presentation the basics of  the quark-diquark model with the trigonometric
Rosen-Morse potential, closely following Refs.~\cite{TQC}, \cite{PRD10}. The model  can be formulated both in three dimensional flat space, and
on the three dimensional hypersphere, $S^3$. 
The flat-space interaction under consideration  is given as,
\begin{equation}
{\mathcal V}_{RMt}\left(\frac{r}{d} \right)
=\frac{1}{d^2}\ell (\ell +1) \csc^2\left(\frac{r}{d}\right) -2\frac{b}{d^2}\cot \left(\frac{r}{d}\right), \quad b=\frac{2\mu dG}{\hbar^2}.
\label{RMT_pt}
\end{equation}
In flat space, $d$ is only a matching length parameter, while  for  a ${\mathcal V}_{RMt}$  placed on a compact hyperspherical surface, $S^3_R$, it
acquires meaning of the constant hyper-radius $R=d$. On $S^3_R$, the $\csc^2$ term is absorbed into the Laplace-Beltrami operator and thereby
into the free motion. The remaining cotangent term, is a harmonic function on $S^3_R$, a reason for which 
it is frequently referred  to as  ``curved'' Coulomb.  
The Hamiltonian, ${\mathcal H}_{RMt}$,  describing in the equivalent one dimensional Schr\"odinger equation such a motion, 
can be shaped as a  Casimir invariant, ${\widetilde {\mathcal K}}$, of a non-trivial representation of the $so(4)$ sub-algebra of the 
conformal  $so(2,4)$,  a result reported in \cite{PalKi}.
{}For the ground state, this representation is specifically simple and is given by,
\begin{eqnarray}
{\mathcal H}_{RMt}={\widetilde {\mathcal K}} -b^2, &\quad&
{\widetilde {\mathcal K}}=\sin^{-1}\chi e^{-b\chi}\, {\mathcal K}e^{b\chi }\, \sin\chi ,
\end{eqnarray}
where ${\mathcal K}$ is the regular geometric $so(4)$ Casimir operator on $S^3_R$,  $\chi$ is the second polar angle 
in the parametrization of the hypersphere.
Important, the first terms in the Taylor  series expansion of ${\mathcal V}_{RMt}$,
coincide with  the Cornell potential, predicted by Lattice QCD,  in the presence of a centrifugal barrier, i.e. with 
\begin{eqnarray}
{\mathcal V}_{RMt}\left( \frac{r}{d}\right)&\approx & \frac{\ell(\ell+1)}{r^2} -\frac{4\mu G}{\hbar^2 r} +\frac{4\mu G}{3\hbar^2 d^3}r,
\label{Cornell}\\
\lim_{d\to \infty} {\mathcal V}_{RMt}\left( \frac{{r}}{d}\right)&\longrightarrow &\frac{\ell(\ell+1)}{r^2} -\frac{4\mu G}{\hbar^2 r}.
\label{Cornell_flat}
\end{eqnarray}
In case ${\mathcal V}_{RMt}$ where to be considered on the hypersphere, 
the radial distance, $r$,  would become $\stackrel{\frown}{r}$, the arc of an $S^3_R$  geodesic and  $d$ would take the place of the hyper-radius, $R$.  
As visible from eq.~(\ref{Cornell_flat}), in the flat space limit,  ${\mathcal V}_{RMt}$ approaches the inverse distance potential,
i.e. again a conformal interaction. In this manner, the trigonometric Rosen-Morse potential is conformally invariant in both curved- 
and flat spaces.  This is different from the case of the Higgs oscillator potential problem  on the hyperbolic plane,
which acquires conformal symmetry exclusively in the flat space limit. 

\begin{itemize}
\item In effect, the trigonometric Rosen-Morse potential provides a  through and through  conformal interaction in the infrared and is 
a manifest example for the possibility of realizing conformal symmetry in the presence of two scales-- 
the strength of the cotangent term, $G$,  and the length parameter $d$, equivalent to the hyperspherical radius $R$. 
The last scale allows for the introduction of temperature as the inverse hyper-radius, $T=1/R$ \cite{Tim}.
\end{itemize}
 Though the potential under discussion captures essential relativistic features brought about by the local isomorphism between $so(4)$ and the Lorentz algebra $so(1,3)$, 
such as the dimensionality of the irreducible representation functions,
in operating on a compact manifold, and in having infinitely many bound states,
it is at the same time  closely tied to  near rest-frame physics. 
In effect, this potential gives rise to a conformal spectrum of  hydrogen-like degeneracy patterns, 
according to
\begin{equation}
\frac{2\mu E^{RMt}_{n\ell}}{\hbar^2} =-\frac{4\mu^2 G^2}{\hbar^4\left(n+\ell +1\right)^2} + \frac{1}{d^2}(n+\ell +1)^2
\stackrel{d\to \infty}{\longrightarrow}-\frac{4\mu^2 G^2}{\hbar^4\left(n+\ell +1\right)^2},
\label{RMt_enr}
\end{equation}
again in the units of fm$^{-2}$ used through the paper. The latter equation makes manifest 
one more difference between the trigonometric quark model in the infrared and the
hyperbolic one in the ultraviolet. The issue is that while in the ultraviolet setup  the physical spectrum is explained by the small hyperbolic angle limit  
to P\"oschl-Teller II in (\ref{bo2}), in the infrared it requires the full angular range of the trigonometric Rosen-Morse potential. 
The latter spectrum  matches well the observed degeneracies in both the non-strange baryon- and meson excitations in the 1500 MeV to 2500 MeV range,
a phenomenon whose appearance is justified by the opening of the conformal window \cite{Andre} in the infrared. Moreover, 
upon being employed as a gauge potential in the Klein-Gordon scale equation on the three dimensions hypersphere, $S^3$,
it furthermore provided a realistic description of the 
$P_{2I, 1}$--$S_{2I,1}$ orderings  through the entire known $N$ and $\Delta$ spectra as a kinematic splitting effect \cite{PRD10}. 
Within this framework, the respective protob charge-electric form factor,  $G_E^p(Q^2)$,  has been predicted as,
\begin{eqnarray}
G_E^p(Q^2)&=&\frac{b(b^2+1)}{Qd}
\tan^{-1} 
\frac{16bdQ}
{(Qd)^4 +4(2b^2-1)(Qd)^2 +16b^2(b^2+1)  },
\label{RMt_FF}
\end{eqnarray}
with $b$ from (\ref{RMT_pt}). The $b$ and $d$ values are those given in \cite{TQC}.

\vspace{0.53cm}
\noindent
{\bf Acknowledgment}: We thank Cliffor Compean for the critical reading of the manuscript and valuable comments.

\end{document}